\documentclass[12pt]{article}
\usepackage{graphicx}
\usepackage{latexsym}
\def\a{\alpha}
\def\b{\beta}
\def\g{\gamma}
\def\d{\delta}

\def\l{\lambda}

\def\m{\mu}
\def\n{\nu}

\def\o{\omega}
\def\s{\sigma}

\def\be{\begin{equation}}
\def\ee{\end{equation}}
\def\beq{\begin{eqnarray}}
\def\eeq{\end{eqnarray}}

\newsavebox{\uuunit}
\sbox{\uuunit}
    {\setlength{\unitlength}{0.825em}
     \begin{picture}(0.6,0.7)
        \thinlines
        \put(0,0){\line(1,0){0.5}}
        \put(0.15,0){\line(0,1){0.7}}
        \put(0.35,0){\line(0,1){0.8}}
       \multiput(0.3,0.8)(-0.04,-0.02){12}{\rule{0.5pt}{0.5pt}}
     \end {picture}}
\newcommand {\unity}{\mathord{\!\usebox{\uuunit}}}
\newcommand{\bqn}{\begin{eqnarray}}\newcommand{\eqn}{\end{eqnarray}}
\newtheorem{theorem}{Theorem}[subsection]

\begin{document}

\begin{titlepage}
\begin{flushright}
hep-th/0206125\\
IHES/P/02/43 \\
ULB-TH-02/18 \\
\end{flushright}

\begin{centering}

{\large {\bf Einstein billiards and overextensions of
finite-dimensional simple Lie algebras}}

\vspace{.5cm}

Thibault Damour$^{a}$, Sophie de Buyl$^b$, Marc
Henneaux$^{b,c}$ \\
and \\
Christiane Schomblond$^{b}$ \\
\vspace{.7cm} {\small $^a$ Institut des Hautes Etudes
Scientifiques, 35, Route de
Chartres,  F-91440 Bures-sur-Yvette, France \\
\vspace{.2cm} $^b$ Physique Th\'eorique et Math\'ematique,
Universit\'e Libre
de Bruxelles,  C.P. 231, B-1050, Bruxelles, Belgium      \\
\vspace{.2cm} $^c$ Centro de Estudios Cient\'{\i}ficos, Casilla
1469, Valdivia, Chile}

\vspace{.5cm}

\end{centering}

\begin{abstract}
In recent papers, it has been shown that (i) the dynamics of
theories involving gravity can be described, in the vicinity of a
spacelike singularity, as a billiard motion in a region of
hyperbolic space bounded by hyperplanes; and (ii) that the
relevant billiard has remarkable symmetry properties in the case
of pure gravity in $d+1$ spacetime dimensions, or supergravity
theories in $10$ or $11$ spacetime dimensions, for which it turns
out to be the fundamental Weyl chamber of the Kac-Moody algebras
$AE_d$, $E_{10}$, $BE_{10}$ or $DE_{10}$ (depending on the model).
We analyse in this paper the billiards associated to other
theories containing gravity, whose toroidal reduction to three
dimensions involves coset models $G/H$ (with $G$ maximally non
compact). We show that in each case, the billiard is the
fundamental Weyl chamber of the (indefinite) Kac-Moody
``overextension'' (or ``canonical Lorentzian extension'') of the
finite-dimensional Lie algebra that appears in the toroidal
compactification to $3$ spacetime dimensions. A remarkable feature
of the billiard properties, however, is that they do not depend on
the spacetime dimension in which the theory is analyzed and hence
are rather robust, while the symmetry algebra that emerges in the
toroidal dimensional reduction is dimension-dependent.

\end{abstract}

\vfill
\end{titlepage}

\section{Introduction}
\setcounter{equation}{0} \setcounter{theorem}{0}
\setcounter{lemma}{0}

\subsection{Einstein billiards}
This paper is devoted to a further investigation of the remarkable
regularity properties of the billiards in hyperbolic space that
control the asymptotic dynamics of theories involving gravity in
the vicinity of a spacelike singularity.

Motivated by the work of Belinskii, Khalatnikov and Lifshitz
\cite{BKL}, it was recently found that the dynamics of the
Einstein-dilaton-$p$-form system in any number of spacetime
dimensions can be described, near a spacelike singularity, as a
billiard motion in a region of hyperbolic space \cite{DH1,DH3}.
The dimension of the billiard, which will be called ``Einstein
billiard", and its precise shape depend on the theory at
hand\footnote{For four-dimensional vacuum gravity, one recovers
the billiards described in \cite{Chitre,Misnerb}.}. The emergence
of hyperbolic geometry is related to the fact that the
``supermetric" \cite{DeWitt0} in the space of spatial metrics has
Lorentzian signature. [A detailed derivation of the billiards will
be given in \cite{DHN}.]

It was further realized in \cite{DH3} that the billiards
associated (i) with the low energy bosonic sectors of
$11$-dimensional supergravity, or of types $II_A$ and $II_B$
supergravities in $10$-dimensions, or with $10$-dimensional type
$I$ supergravities (ii) with or (iii) without a vector multiplet,
could be identified with the fundamental Weyl chamber of the
Kac-Moody algebra $E_{10}$ (case (i)), $BE_{10}$ (case (ii)) or
$DE_{10}$ (case (iii)), which are all of indefinite type. These
striking symmetry properties of the billiards relevant to the
asymptotic dynamics hold also for pure gravity in any number of
spacetime dimensions $D = d+1$, for which the billiard turns out
to be the fundamental Weyl chamber of $AE_d$ \cite{DHJN}.  The
billiard dynamics is chaotic when the Kac-Moody algebra is
hyperbolic \cite{DHJN}, which is the case for $E_{10}$, $BE_{10}$,
$DE_{10}$ and $AE_d$ with $d \leq 9$.  Vacuum gravity is no longer
chaotic in spacetime dimensions $\geq 11$, as observed previously
in \cite{DHS}.

The purpose of this paper is to analyze in the light of these
results other theories involving gravity, which have been found to
be of interest in the past.  These theories were introduced and
classified by Breitenlohner, Maison and Gibbons \cite{BGM}. The
motivation of these authors in formulating such general classes of
theories was to generalize the existence and uniqueness theorems
known for black holes in Einstein-Maxwell theory. These theories,
initially formulated in Ref. \cite{BGM} as four-dimensional
theories, have the property that, when reduced to $3$ spacetime
dimensions, they describe three-dimensional gravity coupled to a
coset model $G/H$, with a finite-dimensional symmetry group $G$.
In many cases the initial four-dimensional theories can themselves
be obtained from the dimensional reduction of higher-dimensional
theories. On then says that they can be ``oxidized'' to higher
dimensions. Their oxidation to higher dimensions has been
systematically explored in \cite{CJLP} and we shall closely follow
the formulation given in that paper.  Some, but not all, of these
theories are related to supergravity.  As in \cite{CJLP}, we
concentrate on the cases where the numerator group $G$ of the
three-dimensional coset is maximally non-compact (``normal" or
``split" real form of the corresponding Lie algebra ${\cal G}$).
We compute in each case the relevant billiard and find the
remarkable result that the billiard is always the fundamental Weyl
chamber of the ``overextension'' ${\cal G}^{\wedge\wedge}$ of the
finite-dimensional Lie algebra ${\cal G}$ that appears in the
toroidal compactification to three spacetime dimensions. We shall
recall in Section \ref{overex} the definition of the overextension
(or canonical Lorentzian extension) of a finite-dimensional Lie
algebra ${\cal G}$, i.e. the canonical association to any ${\cal
G}$ (of rank $r$) of a certain indefinite Kac-Moody algebra ${\cal
G}^{\wedge \wedge}$ with rank $ r^{\wedge \wedge} = r + 2$. Let us
only note beforehand that the Kac-Moody algebras ${\cal
G}^{\wedge\wedge}$ are symmetrizable and that the metric in the
Cartan subalgebra has Lorentzian signature, reflecting the
signature of the supermetric for gravity.

Our main result, the sytematic appearance of the overextended
symmetry algebra ${\cal G}^{\wedge \wedge}$, is a vast
generalization of what was found to hold in the particular cases
of eleven- and ten-dimensional supergravity theories \cite{DH3} or
pure gravity \cite{DHJN}. We note also that, if we view (as is
explained in \cite{DH2,DH3,DHN}) the cosmological billiards as a
kind of ``one-dimensional reduction'' this result confirms the
conjecture made long ago by Julia \cite{Julia} concerning the
systematic appearance of an ``affine extension'' ${\cal
G}^{\wedge}$ in the reduction from $3$ to $2$ dimensions, to be
followed by the appearance of an overextension ${\cal G}^{\wedge
\wedge}$ in the further ``reduction to one dimension''  (in a
sense which remained to be defined).

Contrary to what happens when considering (toroidal)
dimensional reductions of a given higher-dimensional theory
(for which the symmetry algebra explicitly
depends on the dimension to which the model is reduced, i.e. on the
height on the oxidation sequence),
our billiard calculations can be performed in
any number of dimensions, and always lead to the same result:
namely the Weyl group of some  Kac-Moody algebra. In practice, however,
there are two preferred dimensions for doing the calculation:
either the upper end of the oxidation sequence (i.e. the highest
possible dimension; which corresponds to the most economical formulation
of the theory), or its lower end (i.e. in $3$ spacetime dimension;
where the ``initial'' symmetry group ${\cal G}$ is present).
In fact, both ways of doing the calculation have their respective
merits. The highest-dimension calculation is technically the
most straightforward. For this reason, we shall present it in detail
in the main body of the paper in which we prove, by a case by case analysis,
our main result. However, we shall present in our concluding section
how a systematic reasoning done directly in the lowest dimension
($D=3$) can elegantly detect the origin of the systematic
appearance of an overextension.

\subsection{Organization of the paper and conventions}

Our paper is organized as follows.  In the next section, we
briefly recall how the billiards emerge from the dynamics in the
vicinity of a spacelike singularity for the general system
described by the Lagrangian \beq {\cal L}_D &=& ^{(D)}R\star \unity -
\sum_\a \star d\phi^\a \wedge d\phi^\a \nonumber\\ && -
\frac{1}{2} \sum_p  e^{\l^{(p)} (\phi)} \star F^{(p+1)}\wedge
F^{(p+1)}, \; \; \; D \geq 3    \label{keyaction} \eeq where
$\l^{(p)}(\phi) \equiv \sum_\a \l^{(p)}_\a \phi^\a$, and where we
have chosen units such that $16 \pi G = 1$. We allow $N$
``dilatons" $\phi^\a$ ($\a = 1,2, \cdots, N$) and normalize their
kinetic terms with a weight 1 with respect to the Ricci scalar.
The Einstein metric $g_{\m \n}$ has Lorentz signature $(-, +,
\cdots, +)$ and is used to lower or raise the indices. Its
determinant is $\!^{(D)}g$. The integer $p \geq 0$ labels the
various $p$-forms $A^{(p)}$ present in the theory, with field
strengths $F^{(p+1)} \equiv dA^{(p)}$,
 \be F^{(p+1)}_{\m_1 \cdots \m_{p+1}} =
\partial_{\m_1} A^{(p)}_{\m_2 \cdots \m_{p+1}} \pm p \hbox{
permutations }.  \ee In fact, the field strength can be modified
by additional coupling terms of Yang-Mills or Chapline-Manton type
(e.g., $F_C^{(3)} = dC^{(2)} - C^{(0)} dB^{(2)}$ for two $2$-forms
$C^{(2)}$ and $B^{(2)}$ and a $0$-form $C^{(0)}$, as it occurs in
ten-dimensional type IIB supergravity).  One can also add
Chern-Simons terms.  Although neither of these modifies the
billiard analysis \cite{DH1,DHN}, for the sake of completeness, we
shall write them down explicitly below for the various theories
under consideration (note that these terms are important for the
symmetry group analysis of \cite{CJLP}). If there are several
$p$-form gauge fields with the same form degree $p$, we use
different letters $A^{(p)}$, $B^{(p)}$, $C^{(p)}$, ... to
distinguish them, as we just did. The real parameters
$\l^{(p)}_\a$ measure the strength of the coupling to the
``dilatons'' $\phi^\a$.
 When $p=0$, we assume that $\l^{(0)}_\a \not=0$ (for at
least one $\a$), i.e., we assume that we have collected all the
dilatons in the $\phi^\a$'s. We shall also recall in section
\ref{recipes}
the rules for computing the billiard walls as well as their angles
and normalization; these rules have been given in \cite{DH1,DH3}
and are explicitly written here in the case of many dilatons.
We also verify that the billiard is invariant under toroidal
dimensional reduction.

In section \ref{general}, we show that, if all the dilatons are
absent, and if there are only {\it proper}  $p$-forms,\footnote{
We define ``proper $p$-forms'' as the $p$-forms with $0<p<D-2$.
Indeed, $p=0$ corresponds to a scalar, $p = D-2$ is dual to a
scalar and $p = D-1$ corresponds to the addition of a cosmological
constant.} the billiard has always a finite volume. Hence, in that
case, the system is always chaotic according to general results on
hyperbolic billiards \cite{billiards}.  We also emphasize that
although a billiard description of the asymptotic dynamics (as one
approaches a spacelike singularity) always exists, it is only for
special spacetime dimensions, menus of $p$-form and dilaton
couplings that the billiard in question has notable regularity
properties.  In particular, the billiard associated with the
gravity+$3$-form system is a Coxeter polyhedron only in $11$
spacetime dimensions.

In section \ref{overex}, we recall the definition of
overextensions of finite-dimensional Lie algebras
\cite{GO,Julia,Kac}.  We consider next the models associated with
the classical groups (section \ref{classical}) and the exceptional
groups (section \ref{exceptional}) and show that in each case, the
billiard is the fundamental Weyl chamber of the overextension of
the corresponding Lie algebra. Finally, we close our paper with
some conclusions and with a streamlined argument explaining the
origin of the systematic appearance of the overextensions.

\section{Recipes for constructing billiards}
\label{recipes} \setcounter{equation}{0} \setcounter{theorem}{0}
\setcounter{lemma}{0}\subsection{Billiard walls} We recall the
rules for constructing the billiards associated with the system
(\ref{keyaction}) \cite{DH1,DH2,DHN}.  As one goes toward a
spacelike singularity, the various spatial points effectively
decouple \cite{BKL}.  At each spatial point, the degrees of
freedom that carry the essential dynamics are the logarithms $\b^i
\equiv - \ln a_i$ of the scale factors $a_i$ along a set of
(special) independent spatial directions ($ds^2 = -dt^2 +\sum_i
a_i^2(t,x)(\o^i)^2$) and the dilatons $\phi^\a$. [See \cite{DHN}
for the definition of the special vielbeins $\o^i = e^i_j(t,x) d
x^j$ leading to a simple billiard dynamics, as $ t \to 0$, for the
scale factors $a_i(t,x)$ after elimination of the off-diagonal
components described by $ e^i_j(t,x)$.] We denote collectively
these variables by $\b^\m$ ($\m = 1, \cdots, M$) and shall loosely
call all of them ``scale factors". The total number of ``scale
factors'' is therefore $ M = d + N$, where $d$ is the spatial
dimension and $N$ the number of dilatons. These variables are
constrained (by the dynamics) to lie in a convex cone (``the wall
cone") defined by a set of linear inequalities of the form \be
w_A(\beta) \equiv w_{A \m} \b^\m \geq 0 \; \; \; A= 1, \cdots Q.
\ee Here $Q$ is the number of  walls, which depends on the system.
The $w_{A \m}$ are the ``wall forms" and the hyperplanes
$w_A(\beta) = 0$ are the ``walls". Note that the {\it wall cone}
defined by the inequalities above does not depend on the
normalization of the wall forms $w_{A \m}$ (i.e. the same cone
would be defined by considering wall forms  $ \l_A w_{A \m}$ with
$\l_A$ being any positive factor). However, the billiard dynamics
that we consider is always characterized by a preferred set of
wall forms, with a well-defined normalization. In fact, as shown
in \cite{DH3,DHN} our asymptotic dynamics is richer than a pure
billiard dynamics with infinitely sharp walls. It is a
(generalized) Toda dynamics, where each wall is a ``Toda wall'',
i.e. corresponds to an exponential term $ c_A^2 \exp( - 2 w_A(\b)
)$ in the Hamiltonian. The exponent appearing in these potential
terms uniquely define our normalization for the wall forms.

Between two collisions against the walls, the scale factors move
along a null straight line of the $M-$dimensional (where we recall
that $ M = d + N$) flat metric \be dS^2 = G_{\m \n}
\b^\m \b^\n = \sum_i (d \b^i)^2 - (\sum_i d \b^i)^2 + \sum_\a (d
\phi^\a)^2 \label{dewitt}\ee  This metric has Lorentzian signature
$(-,+,+, \cdots, +)$.  The inverse metric reads \be \label{Gmn} G^{\m
\n}
\partial_\m f
\partial_\n f = \sum_i (\partial_i f)^2 - \frac{1}{d-1}(\sum_i
\partial_i f)^2 + \sum_\a (\partial_\a f)^2 \ee
for any function $f(\b^i, \phi^\a)$.  The future light cone of the
metric is defined by  $G_{\m \n} \b^\m \b^\n < 0$ and $\sum_i \b^i >0$.

The list of the walls that must be a priori considered is as follows:
\begin{enumerate}
\item {\bf Symmetry walls} \cite{DH3,DHN}. These arise from the
off-diagonal components of the metric and are cleanly derived
using the Iwasawa decomposition of the spatial metric \cite{DHN}.
They read explicitly: \be w_{ij}^S (\b) = \b^j - \b^i, \; \; \; \;
\; i<j \label{sym}\ee \item {\bf Gravitational (or curvature)
walls} \cite{DH1,DHS}. These arise from the spatial curvature and
are given by \be w^G_{i;jk} (\b)= 2 \b^i + \sum_{l \not = i,j,k}
\b^\ell, \; \; \; \; \; i \not=j, \; i \not= k, \; j \not= k
\label{Curv}\ee These walls are absent for $D=3$, i.e., there is
no gravitational wall to be considered in that dimension. [There
are, in all dimensions, subdominant gravitational walls of the
form $w_i^G(\b) = \sum_{\ell \not=i} \b^\ell$. These walls are
lightlike \cite{DHN} and do not affect the asymptotic dynamics of
the scale factors; this is why they have not been listed.  Note
that we assume at least one matter field  when $D=3$ (which can be
assumed to be a scalar field by dualization), since otherwise,
there is no local degree of freedom and the only interesting
dynamics is in the global degrees of freedom, not discussed
here.]\item {\bf Electric walls} \cite{DH1}. These arise from the
electric energy density. For a $p$-form, the walls read \be w^{E
\, (p)}_{i_1 \cdots i_p}(\b) = \b^{i_1} + \cdots + \b^{i_p} +
\frac{1}{2} \sum_\a \l^{(p)}_\a \phi^\a , \; \; \; \; \; i_1 <
\cdots < i_p \label{elec}\ee \item {\bf Magnetic walls}
\cite{DH1}. These arise from the magnetic energy density.  For a
$p$-form, they read \be w^{M \, (p)}_{i_1 \cdots i_{d-p-1}} (\b) =
\b^{i_1} + \cdots + \b^{i_{d-p-1}} -  \frac{1}{2} \sum_\a
\l^{(p)}_\a \phi^\a , \; i_1 < \cdots < i_{d-p-1}\label{magn} \ee
\end{enumerate}

Not all these walls are relevant since the inequalities $w^S_{ij}
\geq 0$, $w^A_{i;jk} \geq 0$, $w^E_{i_1 \cdots i_p} \geq 0$ and
$w^M_{i_1 \cdots i_{d-p-1}} \geq 0$ follow from the simpler subset
\be \b^1 \leq \b^2 \cdots \leq \b^d, \; \; w^G_{1;2 \, 3} \geq 0,
\; \; w^E_{1 \cdots p} \geq 0, \; \; w^M_{1 \cdots \, d-p-1} \geq
0 \label{subset} \ee This subset might still be redundant; e.g.,
in the presence of a proper $p$-form (i.e. with $0<p<D-2$),
the gravitational
wall forms can be written as sums of one electric and one magnetic
wall form. So, once the list of all walls (\ref{subset}) has been
written down for a given theory, the first task is to determine
which among these walls are the ``dominant'' ones, i.e. the minimal
set of walls which suffice to define the billiard.

It is easy to check that all the walls are {\it timelike}, i.e.,
that their normals are spacelike, $w_A \cdot w_A >0$
\cite{DH1,DH3}. In fact, the wall cone contains the future
pointing timelike vector $\b^1 = \b^2 = \cdots = \b^d = \alpha
>0 $, $\; \phi^1 = \phi^2 = \cdots = \phi^N = 0$.  The wall cone
has therefore a non trivial intersection with the future light
cone.  Two cases can arise \cite{DHJN}
\begin{enumerate}
\item \label{case1} The wall cone is entirely contained in the
light cone; in this case, the point representing the system in the
space of the scale factors (``billiard ball") never stops hitting
the walls. \item \label{case2} The wall cone is not entirely
contained in the light cone and contains thus some spacelike
straight lines through the origin; in this case, the billiard ball
will generically make a finite number of collisions with the walls
and then go on freely forever, the directions of escape being
parallel to the lightlike straight lines through the origin
contained in the wall cone.
\end{enumerate}

One can project the dynamics on the upper sheet of the unit
hyperboloid $G_{\m \n} \b^\m \b^\n = -1$ \cite{DH3}. This
hyperboloid can be identified with the hyperbolic space ${\cal
H}_{d+N-1}$. The walls define hyperplanes in ${\cal H}_{d+N-1}$.
The interior region bounded by these hyperplanes is the billiard.
The billiard is therefore the radial projection of the wall cone
onto the unit hyperboloid. In case \ref{case1}, the billiard has finite
volume; in case \ref{case2}, it has infinite volume.

\subsection{Dimensional reduction - Dualization} The billiard is
defined by the action (\ref{keyaction}) and by nothing else. One
of its crucial properties is that it is invariant under toroidal
compactification down to any spacetime dimension $\geq 3$ below
the original spacetime dimension.  The walls simply change name,
but the billiard remains the same. Indeed, consider toroidal
compactification of just one dimension (the general case is
obtained by iteration) of (\ref{keyaction}) with only one
$p$-form.  The $D$-dimensional metric is related to the
$(D-1)$-dimensional (Einstein)
metric $\hat g_{\m \n}$, the Kaluza-Klein (KK)
vector $\hat{ \cal A}_{ \m}$ and the additional dilaton $\hat
\varphi$ through the formulas \begin{eqnarray} g_{11} &=& e^{-2
(d-2) \g \hat \varphi}, \\ g_{1 \m} &=& e^{-2(d-2) \g \hat
\varphi} \hat {\cal A}_\m , \; \; \; \; \; \m = 0,2,3, \cdots, d \\
g_{\m \n} &=& e^{2 \g \hat \varphi} \hat g_{\m \n} + e^{-2(d-2) \g
\hat \varphi} \hat {\cal A}_\m \hat {\cal A}_\n
\end{eqnarray} where $\gamma =
1/\sqrt{(D-2)(D-3)}$.  The $(D-1)$-form of the action reads \beq
&& {\cal L}_{D-1} = ^{(D-1)}\hat R\star \unity - \star
d\hat\varphi\wedge d\hat\varphi -\frac{1}{2}
e^{-2(D-2)\gamma\hat\varphi}\star\hat{\cal F}\wedge\hat{\cal F} -
\sum_\a \star d\hat\phi^\a \wedge d\hat\phi^\a \nonumber\\ && -
\frac{1}{2}
 e^{\l^{(p)}_\a \hat\phi^\a -2 p\gamma\hat\varphi}
\star \hat F^{(p+1)}\wedge \hat F^{(p+1)} - \frac{1}{2}
 e^{\l^{(p)}_\a \hat\phi^\a +2(D-2- p)\gamma \hat\varphi}
\star \hat F^{(p)}\wedge \hat F^{(p)}    \eeq where $\hat{\cal F}$
is the field strength of the K-K vector potential and where $\hat
F^{(p+1)}$ and $\hat F^{(p)}$ denote the field strengths of the
$p$-form and the $(p-1)$-form originating from the $p$-form in $D$
dimensions.

Let $\hat \b^a$ ($a = 2,\cdots,d$) be the scale factors in $D-1$
spacetime dimensions. The relationship between $\hat \varphi$,
$\hat \b^a$ and the original scale factors $\b^i$ ($i= 1,a)$ in
$D$ spacetime dimensions is \be \b^a = \hat \b^a - \g \hat
\varphi, \; \; \; \b^1 = (d-2) \g \hat \varphi. \ee Given this
relationship, one easily verifies by mere substitution that
\begin{itemize}
\item The $d(d-1)/2$ symmetry walls (\ref{sym}) give rise to
$(d-1)(d-2)/2$ symmetry walls in $(D-1)$ dimensions \be \hat w
^S_{ab} = \hat \beta^b - \hat \beta^a, \,\, a<b\ee and to the
$(d-1)$ electric walls of the K-K $1$-form: \be \hat w^{E,KK}_a =
\hat\beta^a - (d-1)\gamma\hat\varphi.\ee \item The $d!/ p! (d-p)!$
electric walls (\ref{elec}) give rise to the $(d-1)!/p!(d-1-p)!$
electric walls of the $p$-form: \be \hat w^{E,p}_{a_1\cdots a_p} =
\hat\beta^{a_1}+\cdots+\hat\beta^{a_p} +
\frac{\lambda_\alpha^{(p)}}{2}\hat\phi^\alpha
-p\gamma\hat\varphi\ee and to the $(d-1)!/(p-1)!(d-p)!$ electric
walls of the $(p-1)$-form: \be  \hat w^{E,p-1}_{a_1\cdots a_{p-1}}
= \hat\beta^{a_1}+\cdots+\hat\beta^{a_{p-1}} +
\frac{\lambda_\alpha^{(p)}}{2}\hat\phi^\alpha
+(d-1-p)\gamma\hat\varphi\ee \item The $d!/(p+1)!(d-1-p)!$
magnetic walls (\ref{magn}) give rise to the
$(d-1)!/(p+1)!(d-2-p)!$ magnetic walls of the $p$-form: \be \hat
w^{M,p}_{a_1\cdots a_{d-2-p}} = \hat\beta^{a_1}+\cdots +
\hat\beta^{a_{d-2-p}} -
\frac{\lambda_\alpha^{(p)}}{2}\hat\phi^\alpha
+p\gamma\hat\varphi\ee and to the $(d-1)!/p!(d-1-p)!$ magnetic
walls of the $(p-1)$-form: \be \hat w^{M,p-1}_{a_1\cdots
a_{d-1-p}} = \hat\beta^{a_1}+\cdots + \hat\beta^{a_{d-1-p}} -
\frac{\lambda_\alpha^{(p)}}{2}\hat\phi^\alpha
-(d-1-p)\gamma\hat\varphi\ee
\end{itemize}

The situation with the gravitational walls is slightly more
subtle.  Indeed, while it is easy to see that the curvature walls
(\ref{Curv}) with $(i,j,k) = (a,b,c)$ are just the curvature walls
in $(D-1)$ dimensions: \be w^{G}_{a;bc} = \hat w^{G}_{a;bc} =
2\hat\beta^a + \sum_{g\ne a,b,c}\hat\beta^g,\;\; a\ne b, a\ne c,
b\ne c\ee  and that the curvature walls  (\ref{Curv}) with
$(i,j,k) = (1,a,b)$ are just the $(d-1)(d-2)/2$ magnetic walls of
the K-K $1$-form (with $\{a_1,...,a_{d-3}\}$ in the complementary
subset to $\{a,b\}$): \be \hat w^{M,KK}_{a_1...a_{d-3}} =
\hat\beta^{a_1}+\cdots + \hat\beta^{a_{d-3}} +
(d-1)\gamma\hat\varphi,\ee one finds that the original
gravitational walls $w^{G}_{a;1b}$ are absent in the
$(D-1)$-dimensional theory.  This is because the corresponding
walls are multiplied, in the $D$-dimensional theory, by the
(square of the) structure constants $C^a_{\; \; 1 \, b}$
\cite{DHN}, which are zero under the dimensional reduction
assumption that the fields do not depend on the coordinate $x^1$.
The fact that the walls $w^{G}_{a;1b}$ are absent in the
$(D-1)$-dimensional theory is, however, not a problem because
these walls are always subdominant: the corresponding wall forms
can be expressed as linear combinations with positive (integer)
coefficients of the other gravitational wall forms and the
symmetry wall forms.

Another important -- and rather obvious -- property of the set of
walls is that it is invariant under electric-magnetic duality
transformations, in which one replaces a $p$-form by a
$(D-p-2)$-form such that the curvatures are dual to one another.
This changes the sign of the dilaton coupling and exchanges
electric and magnetic walls.  Note that the squared norm of the
electric and magnetic walls associated with the same $p$-form are
equal.

\subsection{Coxeter polyhedra - Kac-Moody billiards}
A convex polyhedron in ${\cal H}_{d+N-1}$ is a finite intersection
of half spaces of ${\cal H}_{d+N-1}$ with a non-empty interior;
our billiards are thus convex polyhedra in ${\cal H}_{d+N-1}$. A
{\it Coxeter polyhedron} is a polyhedron such that the dihedral
angles between adjacent faces are integer submultiples of $\pi$
(i.e., of the form $\pi /k$ where $k$ is an integer $\geq 2$).
For a Coxeter polyhedron, the group generated by the reflections
in the faces is a discrete subgroup of the isometry group of
${\cal H}_{d+N-1}$ (Poincar\'e theorem). It is in fact a {\it
Coxeter group} admitting the following presentation (see
\cite{Vinberg} for details on discrete reflection groups in
hyperbolic spaces): let $s_i$ be the reflection in the face $i$
and let $\pi/m_{ij}$ be the dihedral angle between the faces $i$
and $j$, with $m_{ij} = 1$ if $i=j$, and $m_{ij} = \infty$ if $i$
and $j$ are not adjacent [The hyperplanes of non-adjacent faces of
a Coxeter polyhedron do not intersect \cite{Vinberg}]. Then, the
presentation of the group generated by the $s_i$'s is \be s_i^2 =
1, \; \; \; (s_i s_j)^{m_{ij}} =1 \ee

Although the billiards defined by the action (\ref{keyaction}) are
polyhedra, they fail, in general, to be Coxeter polyhedra.  This
is illustrated in section \ref{general} below.  However, for the
``interesting systems" considered in this paper, the menu of
$p$-forms, the dilaton couplings and the spacetime dimension
conspire to yield polyhedra with faces that meet precisely with
angles that are submultiples of $\pi$. One has actually even more,
namely, for all the systems considered below
 \begin{itemize} \item The billiard is a simplex, i.e., has
exactly $d+ N$ faces. We denote from now on by $w_i = 0$ ($i =
-1,0 \cdots, d+N-2$) the corresponding relevant faces. \item The
dominant wall forms $w_i$ have scalar products such that the
matrix \be A_{i j} = 2\frac{(w_i \vert w_j)}{(w_i\vert
w_i)}\quad\hbox{where}\quad (w_i\vert w_j) \equiv
G^{\mu\nu}w_{i\mu}w_{j\nu} \label{Cmatrix}\ee is a generalized
Cartan matrix, i.e., has non-positive integer off-diagonal
components \cite{Kac} (the other properties are automatic
consequences from the definition: $A_{ii} = 2$ and $A_{ij} =0
\Rightarrow A_{ji} =0$).
\end{itemize}
We shall say that a billiard with these special properties is a
{\it (Lorentzian) Kac-Moody billiard}. By contrast with the
definition of a Coxeter billiard, i.e. motion within a Coxeter
polyhedron, which depended only on dihedral angles, the definition
of a Kac-Moody billiard depends, through the definition
(\ref{Cmatrix}) of the Cartan matrix, on the normalization of the
wall forms. But, as we said, our Toda-like billiard comes equipped
with specially-normalized wall forms.  For a Kac-Moody billiard,
the only possible values of $m_{ij}$ are $1$ ($i=j$), $2$, $3$,
$4$, $6$ and $\infty$.

When $A_{ij}$ is a generalized Cartan matrix, it defines a
Kac-Moody algebra of rank $d+N$, symmetrizable, of indefinite
type, with a scalar product of Lorentzian signature (the scalar
product is just (\ref{dewitt})).  The {\it Kac-Moody billiard}
defined by the inequalities $w_i(\beta) \geq 0$ can then be
identified with the fundamental Weyl chamber of the Kac-Moody
algebra (more precisely, is the radial projection on the unit
hyperboloid of the intersection of the fundamental Weyl chamber
with the  light cone).  Among Lorentzian Kac-Moody algebras, those
whose Dynkin diagram is such that the removal of any node yields
the Dynkin diagram of a Kac-Moody algebra of finite or affine type
are called ``hyperbolic".  Hyperbolic Kac-Moody algebras are
important in the present context: Kac-Moody billiards have finite
volume (and hence are chaotic) if and only if the underlying
algebra is hyperbolic \cite{DHJN}.

\section{Some general results}
\label{general}
 \setcounter{equation}{0} \setcounter{theorem}{0}
\setcounter{lemma}{0}

As expected from the discussion of the previous section,
Kac-Moody billiards, being quite special, must be rather exceptional.
The purpose of this section is to show explicitly that
this is indeed the case.  This is rather clear when
there are dilatons, because the angles between the faces of the
billiards depend on the dilaton couplings, which are continuous
parameters.   By continuously changing these parameters, one
continuously change the angles, which are therefore integer submultiples
of $\pi$ only for special values of the dilaton couplings.

The fact that Kac-Moody billiards are rare is also true in
the absence of dilatons.  To establish this fact, we
first prove the following
\begin{theorem}: Assume that there is no dilaton and that there is
at least one proper $p$-form ($0<p<d-1$).  Then, the billiard has
finite volume (and is thus chaotic) in any number of spacetime
dimensions ($\geq 3$).
\end{theorem}

To demonstrate this theorem, we show that the billiard for
the system gravity + a
single proper $p$-form ($0<p<d-1$) has finite volume.    Since the billiard
for a collection of $p$-forms is the intersection of the
individual billiards
with a single $p$-form involved, the result will follow.  Without
loss of generality, we can assume $2p < d$ because the set
of walls is
invariant under electric-magnetic duality.  The walls are explicitly
\beq
&&\b^1 \leq \b^2 \leq \cdots \leq \b^d \; \;
\; \; \; \; \hbox{(symmetry walls)}  \\
&& \b^1 +  \cdots + \b^p \geq 0 \; \;
\; \; \; \; \hbox{(electric wall)}\\
&& \b^1 + \cdots + \b^{d-p-1} \geq 0 \; \;
\; \; \; \; \hbox{(magnetic wall)}
\end{eqnarray}
Since $d-p-1 \geq p$, the relevant walls are the symmetry and electric walls.
The billiard is defined by
\beq
&&\b^1 \leq \b^2 \leq \cdots \leq \b^d \; \;
\; \; \; \; \hbox{(symmetry walls)}  \label{rele1}\\
&& \b^1 +  \cdots + \b^p \geq 0 \; \;
\; \; \; \; \hbox{(electric wall)}
\label{rele2}
\end{eqnarray}
and is a simplex.  Our claim is that the billiard has finite
volume, or, equivalently, that the wall cone (\ref{rele1}),
(\ref{rele2}) is contained within the light cone.  That is, all
the $d$ intersection edges of any subset of $d-1$ faces are
timelike or null. The edge opposite to the electric wall is $(\b^1
, \b^2, \cdots, \b^d) = (\a,\a ,\cdots ,\a)$ with $\a >0$ and is
timelike since $\a^2 d(1-d)<0$.  The other edges are obtained by
dropping one of the symmetry equalities.  Vectors along these
edges can be taken to be \be (-(p-k), -(p-k), \cdots, -(p-k), k, k
,\cdots , k), \; \; \; \; \; \; k=1,2, \cdots, p \label{first} \ee
(first $k$ components equal to $-(p-k)$, last $d - k$ components
equal to $k$) and \be (0, 0, \cdots, 0, 1,\cdots, 1) \; \; \; \;
\; \; k = p+1, \cdots, d-1 \label{second} \ee (first $k$
components equal to $0$, last $d - k$ components equal to $1$).
The edges (\ref{second}) are easily seen to have squared norm
equal to $(d-k)(1+k-d)$ and are timelike ($k<d-1$) or null
($k=d-1$). The edges (\ref{first}) have squared norm equal to \be
k \big[ - k(p^2 + 2(1-d)p + d(d-1)) + p^2 \big] \label{norm2} \ee
The coefficient of $k$ in the bracket is always strictly negative,
which means that the expression in the bracket is a monotonously
decreasing function of $k$.  Accordingly, if we show that
(\ref{norm2}) is negative for $k=1$, it will be automatically
negative for $k>1$. Now, for $k=1$, the bracket in (\ref{norm2})
reads
$$
- (p^2 + 2(1-d)p + d(d-1)) + p^2
$$
Setting $d = 2p + s$ with $s>0$, this becomes
$$
- s (s + 2p - 1)
$$
which is strictly negative.  Hence, all the edges are timelike or
null and the billiard has finite volume. [We have also learned that
there is only one lightlike edge, and hence, only one vertex at infinity
for a billiard with a single $p$-form and no dilaton.] $\Box$.

Note that the we need in fact only $d \geq 2p$ instead of the
strict inequality $d > 2p$ to achieve this result, since when
$d=2p$, there is just one additional edge on the light cone, but
this still gives a finite-volume billiard. [However, when $d=2p$,
the magnetic wall dominates.] This shows in particular that the
electric walls are sufficient to drive the chaos when  $d \geq
2p$; otherwise chaos is magnetically-driven. The particular cases
$p=1,2,3$ were investigated previously in \cite{DH2}.

A direct application of the above theorem is:
\begin{theorem}: Consider gravity coupled to a collection of proper
$p$-forms, with no dilaton ($0<p<d-1$ for each $p$-form). If the
spacetime dimension $D$ is strictly greater than $11$, $D>11$, the
corresponding billiard cannot be a Kac-Moody billiard.
\end{theorem}

Indeed, if it were, the corresponding Kac-Moody algebra should be
hyperbolic since the system is chaotic.  But there is no
hyperbolic Kac-Moody algebra of rank strictly greater than $10$
\cite{Kac}.
$\Box$.

We now consider the particular case of gravity coupled to a
$3$-form and investigate when the associated billiard is a
Coxeter polyhedron.
\begin{theorem}: Consider the Einstein-$3$-form system in
spacetime dimension $D \geq 6$.  The
billiard associated with this system is a Coxeter polyhedron if
and only if $D=11$.
\end{theorem}
Note the restriction to $D \geq 6$ which corresponds to the case
where the $3$-form is proper: $ p=3 < D-2$. The billiard for this
system is a simplex bounded by the $d-1$ symmetry walls and the
magnetic wall \be \b^1 \geq 0 \label{magn1} \ee for $D=6$, the
magnetic wall \be \b^1 + \b^2 \geq 0 \label{magn2} \ee for $D=7$
and the electric wall \be \b^1 + \b^2 + \b^3 \geq 0 \label{elec1}
\ee for $D \geq 8$.  The magnetic wall (\ref{magn1}) is orthogonal
to all the symmetry walls except $\b^2 - \b^1 = 0$, with which it
makes an angle equal to $\pi/ 5.104301 $, which is not an integer
submultiple of $\pi$. The magnetic wall (\ref{magn2}) is
orthogonal to all the symmetry walls except $\b^3 - \b^2 = 0$,
with which it makes an angle equal to $\pi / 3.614672$, which is
not an integer submultiple of $\pi$.  Finally, the electric wall
(\ref{elec1}) is orthogonal to all the symmetry walls except $\b^4
- \b^3 = 0$, with which it makes an angle $\hat\alpha$ given by
\be \cos\hat\alpha =
\frac{1}{\sqrt{6}}\,\sqrt{1+\frac{3}{d-4}}.\ee This angle is an
integer submultiple of $\pi$ only for $D=11$, in which case it is
equal to $\pi/3$ and the billiard is the Kac-Moody billiard of
$E_{10}$ \cite{DH3}. $\Box$.

{}From this point of view, $D=11$ is thus quite special for the
gravity-$3$-form system, irrespective of any supersymmetry
consideration.  If one did not know about supergravity, but had some
independent reason for considering a $3$-form,
one could
discover from this independent line of insight the peculiar r\^ole
of $D=11$.

\vspace{.2cm}

\noindent {\bf Note:} One checks along similar lines that besides
the case just considered (and its dual $D=11$, $p=6$ formulation),
the only gravity + $p$-form systems (with a single $p$-form) that
lead to Coxeter polyhedra are: (i) gravity + a $1$-form in $D=4$
and $D=5$ spacetime dimensions, discussed in \cite{DHJN}; (ii)
gravity + a $2$-form in $D=5$ and $D=6$ spacetime dimensions (the
first case is dual to the Einstein-Maxwell theory in $5$
dimensions and yield thus $G_2^{\wedge \wedge}$ \cite{DHJN}; the
second case is easily verified to yield $B_3^{\wedge \wedge}$ and
is the oxidation endpoint of the $B_3$-sequence \cite{CJLP}); and
(iii) gravity + a $4$-form in $10$ dimensions, which yields
$E_7^{\wedge \wedge}$ and is one of the natural oxidation
endpoints of the $E_7$-sequence \cite{CJLP} (see below for
detailed computations).

\section{Overextension of finite-dimensional Lie algebras}
\label{overex} \setcounter{equation}{0} \setcounter{theorem}{0}
\setcounter{lemma}{0}

{}Following \cite{GO,Julia,Kac}, we define the overextended
algebra ${\cal G}^{\wedge\wedge}$ of a finite-dimensional Lie
algebra ${\cal G}$ through its roots and generalized Cartan
matrix. The general construction goes as follows. Let $R$ denote
the root space of a finite-dimensional Lie algebra ${\cal G}$ of
rank $d$, and Cartan matrix $A$. Let $\{\alpha_i, i = 1,...,d\}$
be a (particular) set of simple roots, so that \be A_{i j} =
2\frac{(\a_i\vert \a_j)}{(\a_i\vert \a_i)} \ee where $(.\vert .)$
denotes the standard invariant form on $R$, and let $\theta =
\sum_i n_i \a_i$ be the corresponding highest root (with the $n_i$
being some uniquely defined positive integers). Let us normalize
the standard invariant form on $R$ by the condition
$(\theta\vert\theta) = 2$. As a finite-dimensional Lie algebra has
roots of at most two different lengths, this normalization means
that all the ``long roots'' have a squared length 2.

Let $V$ be the $2$-dimensional hyperbolic space with basis
$u_1,u_2$ and bilinear form of signature $(-,+)$ \be (u_1 \vert
u_2) = 1,\quad (u_1\vert u_1) = (u_2\vert u_2) = 0. \ee Then, we
define the {\it canonical Lorentzian extension} of $R$ as the
orthogonal direct sum $V \oplus R$.  It is equipped with an inner
product of Lorentzian signature $(-,+,+,...+)$. [ Note that Kac
\cite{Kac} calls it the canonical hyperbolic extension of $R$, but
because the corresponding Kac-Moody algebra need not be
hyperbolic, we prefer to use the more neutral adjective
``Lorentzian".]
 Defining the following basis of $V\oplus R$ \beq
  && \alpha_{-1} = u_1 + u_2 \\ &&\alpha_0  = -u_2 -
\theta \\&& \alpha_i \hspace{1cm}
(i=1,...,d)
\end{eqnarray} one can form the $(d+2)\times(d+2)$ matrix
$$A_{\mu\nu}^{\wedge \wedge}
 = 2 {(\alpha_\mu \vert \alpha_\nu) \over (\alpha_\mu \vert
\alpha_\mu) } \hspace{1cm} (\mu,\nu=-1,0,...,d).$$  This matrix is
easily verified to be a generalized Cartan matrix. One then defines
the ``overextension" or ``canonical Lorentzian extension" of the original
finite-dimensional Lie algebra $\cal G$ as the Kac-Moody algebra
built on the generalized Cartan matrix  $A^{\wedge \wedge}$.
We shall denote this (infinite-dimensional) overextended Lie algebra
as ${\cal G}^{\wedge\wedge}$.  It has rank $d+2$.

One can also consider the Kac-Moody algebra generated by the
``once-extended'' set of  $(d+1)$ simple roots $(\a_0, \a_i)$. The
corresponding $(d+1)\times(d+1)$ matrix $A^{\wedge}$ is also a
generalized Cartan matrix, albeit a degenerate one ($ \det
A^{\wedge} =0$). This once-extended Kac-Moody algebra, denoted
${\cal G}^{\wedge}$, is called the ``untwisted affine extension''
of ${\cal G}$. The root $\alpha_0$ is called the ``affine root'',
while $\alpha_{-1}$ is called the ``overextended" root. It is
orthogonal to all others except $\alpha_0$:\be (\alpha_{-1}\vert
\alpha_0)=-1,\quad\quad (\alpha_{-1}\vert \alpha_i)=0.\ee

The Dynkin diagram of ${\cal G}^{\wedge\wedge}$ is obtained from
the corresponding diagram of the affine extension ${\cal
G}^\wedge$ by adding a vertex $\alpha_{-1}$ joined with the affine
vertex $\alpha_0$ by a simple link.

\section{Models associated with classical groups}
\label{classical}
 \setcounter{equation}{0} \setcounter{theorem}{0}
\setcounter{lemma}{0}

Let us consider in turn the different $G/H$ models introduced in
\cite{BGM} and further studied in \cite{CJLP}. They are all
naturally labelled by the numerator groups $G$, or equivalently
(knowing that we consider only the maximally non-compact real
forms of the group) by the corresponding (finite-dimensional) Lie
algebras: successively $A_n, B_n, C_n$ and $ D_n$. We shall
consider the models based on exceptional groups in the next
section.

\subsection{The $A_n$ sequence}
The $A_n$ sequence is by definition such that the symmetry group
in $3$ spacetime dimensions is $SL(n+1, R)$. The maximal oxidation
point of the $A_n$ sequence is pure gravity in $D= n+3$ spacetime
dimension \cite{CJLP}.  The Lagrangian is thus \be {\cal L}_D =
R\star \unity, \; \; D = n+3 \ee There are only symmetry and
gravitational walls.  The billiard is defined by the conditions
\be \b^1 \leq \b^2 \cdots \leq \b^d, \; \; 2 \b^1 + \b^2 + \cdots
+ \b^{d-2} \geq 0, \; \; d=n+2 \ee  and is a simplex. The
computation of the Cartan matrix has been done \cite{DHJN}, where
it was found that $A_{ij}$ is just the generalized Cartan matrix
of the overextension $A^{\wedge \wedge}_n$ of the Lie algebra
$A_n$. [$A^{\wedge \wedge}_n$ is also denoted $AE_{(n+2)}$].
 For $d>3$, it reads
$$  A = \left( \begin{array}{cccccccc} 2 & -1 & 0 & 0 & \cdots & 0 &
0 & 0
\\ -1 & 2 & -1 & 0 & \cdots & 0 & 0 & -1 \\ 0 & -1 & 2 & -1 & \cdots &
0 & 0 & 0
\\
\vdots & \vdots & \vdots & \vdots & \cdots & \vdots &\vdots & \vdots\\
0 & 0 & 0 & 0 &\cdots & 2 &-1 & 0\\ 0 & 0 & 0 & 0 &\cdots & -1 & 2 &
-1\\ 0& -1 & 0 & 0 &
\cdots & 0 & -1 & 2
\\
\end{array}
\right)
$$

As explained above, the computation could have been done in any
dimension between $3$ and $n+3$ with identical conclusions. The
corresponding Dynkin diagram is given in figure 1.  The algebra is
hyperbolic for $n<8$.  The extension $AE_9 \equiv A_7^{\wedge
\wedge}$ is the last hyperbolic algebra in the family; $AE_{10}
\equiv A_8^{\wedge \wedge}$ is not hyperbolic.

\subsection{The $B_n$ sequence}
The oxidation endpoint is $D = n+2$ \cite{CJLP}.  The theory comprises
the metric, a
dilaton, a $2$-form, $B$, and a $1$-form, $A$.  The
Lagrangian reads
 \be {\cal L}_D = R\star \unity - \star d\phi\wedge d\phi - \frac{1}{2}
e^{a\sqrt{2}\phi}\star G\wedge G -\frac{1}{2} e^{
a\frac{\sqrt{2}}{2}\phi} \star F\wedge F,
\ee where $a^2 = 8/n$ and \be G = dB + \frac{1}{2}A\wedge
dA,\quad\quad F = dA.\ee In addition to the symmetry and gravitational
walls, one has electric and magnetic walls for the
$2$-form, \be w^{E,B}_{ij} = \beta^i + \beta^j +
a\frac{\sqrt{2}}{2}\phi,\quad\quad w^{M,B}_{i_1...i_{n-2}} =
\beta^{i_1}+\cdots + \beta^{i_{n-2}} - a\frac{\sqrt{2}}{2}\phi
\ee as well as  electric and magnetic walls for the
$1$-form, \be w^{E,A}_i = \beta^i + a\frac{\sqrt{2}}{4}\phi,\quad\quad
w^{M,A}_{i_1\cdots i_{n-1}} = \beta^{i_1}+\cdots + \beta^{i_{n-1}} -
a\frac{\sqrt{2}}{4}\phi .\ee  It is easy to see that the billiard can
be completely defined by the following $n+2$ independent walls
\beq &&
w_{-1} = \beta^{n+1}-\beta^n\\ && w_0 =
\beta^n-\beta^{n-1}\\ && w_1 =\beta^1+...+\beta^{n-2} -
a\frac{\sqrt{2}}{2}\phi \\ && w_2 = \beta^{n-1} -
\beta^{n-2}\\&& w_3 = \beta^{n-2} -\beta^{n-3}\\&&...\\ && w_{n-1} =
\beta^2 -\beta^1
\\ && w_n = \beta^1 + a\frac{\sqrt{2}}{4}\phi.
\end{eqnarray}  A straighforward computation yields the following
generalized Cartan matrix
$$  A = \left( \begin{array}{ccccccccc}
2 & -1 & 0 & 0 & \cdots & 0 & 0 & 0 & 0 \\
-1 & 2 & 0 &-1 & \cdots & 0 & 0 & 0 & 0 \\
0 & 0 & 2 & -1 & \cdots & 0 & 0 & 0 & 0 \\
0 & -1 & -1 & 2 & \cdots & 0 & 0 & 0 & 0 \\
\vdots & \vdots & \vdots & \vdots & \ddots & \vdots & \vdots & \vdots
& \vdots \\
0 & 0 & 0 & -1 & \cdots & 2 & -1 & 0 & 0 \\
0 & 0 & 0 & 0 & \cdots & -1 & 2 & -1 & 0 \\
0 & 0 & 0 & 0 & \cdots & 0 & -1 & 2 & -1 \\
0 & 0 & 0 & 0 & \cdots & 0 & 0 & -2 & 2 \\
  \end{array}
\right)
$$
The corresponding Dynkin diagram is given in figure 1.
One recognizes the
Dynkin diagram of the overextension $B_n^{\wedge \wedge}$
of the finite-dimensional
algebra $B_n$, which is hyperbolic for $n \leq 8$.
For $n=8$, one recovers the Kac-Moody billiard
of $BE_{10}$ given in \cite{DH3}, which controls the
asymptotic dynamics of the low-energy bosonic sectors of
the heterotic and type I superstrings.  The hyperbolic character of
$BE_{10}$ is another way to see that these models are chaotic.

Note that for $n=3$, one can oxidize further to $D=6$ dimensions
\cite{CJLP}.  The corresponding theory is gravity + a $2$-form,
with no dilaton (there is a self-duality condition on the field
strength of the $2$-form, which only removes the degeneracy of the
$2$-form walls and which is not mandatory from the billiard point
of view).

\subsection{The $C_n$ sequence}
The oxidation end point is the $D=4$ theory whose Lagrangian is
given by  \beq {\cal L}_4 &=& R\star \unity - \star d\vec\phi\wedge
d\vec\phi -\frac{1}{2} \sum_{\alpha} e^{2 \vec
\sigma_\alpha.\vec\phi} \star (d\chi^\alpha + \cdots)\wedge
(d\chi^\alpha+ \cdots)
\nonumber \\
&&- \frac{1}{2}\sum_{a=1}^{n-1} e^{\vec e_a.\vec\phi\sqrt{2}}\star
dA^a_{(1)}\wedge dA^a_{(1)}\end{eqnarray} where the ellipsis
complete the ``curvatures" of the $\chi$'s \cite{CJLP}.  The
$(n-1)$ dilatons $\vec \phi = (\phi^1,...,\phi^{n-1})$ are
associated with the Cartan subalgebra of $Sp(2n-2,R)$ and the
$\frac{1}{2}n(n-1)$ axions $\chi^\alpha$ are associated with the
positive roots of $Sp(2n-2,R)$. The fields $A^a_{(1)}$ are
one-forms. The $\vec \sigma_\alpha$ are the positive roots of
$Sp(2n-2,R)$; these can be written in terms of an orthonormalized
basis of $(n-1)$ vectors in Euclidean space ($\vec e_a .\vec e_b =
\d_{a b}$) $\vec e_a$ ($a = 1, \cdots, n-1$) as \be
\vec\sigma_\alpha = \{ \sqrt{2}\vec e_a, \frac{1}{\sqrt{2}} (\vec
e_a \pm \vec e_b),\; a>b\}.\ee The normalization is such that the
long roots have squared length equal to two. The notation $\vec
\sigma_\alpha.\vec\phi$ means
$$ \vec \sigma_\alpha.\vec\phi \equiv \sum_{a=1}^{n-1}\sigma_\alpha^a \phi^a.$$
The simple roots are \be \sqrt{2}\vec
e_1\quad\hbox{and}\quad \frac{1}{\sqrt{2}}
(\vec e_{a+1} - \vec e_a),\; a=1,...,n-2.\ee
The walls are here given by the curvature wall $2\beta^1$, the
symmetry walls \be \beta^3 - \beta^2 \quad\hbox{and}\quad \beta^2
-\beta^1,\ee the electric and magnetic walls of the axions \be
w^{E,\chi^\alpha} =
\vec\sigma_\alpha.\vec\phi\quad\hbox{and}\quad
w^{M,\chi^\alpha}_{ij} = \beta^i + \beta^j -
\vec\sigma_\alpha.\vec\phi \ee and the electric
and magnetic walls of the $1$-forms \be w^{E,A^a}_i = \beta^i +
\frac{\sqrt{2}}{2}\vec e_a.\vec\phi \quad\hbox{and}\quad
w^{M,A^a}_i = \beta^i -\frac{\sqrt{2}}{2}\vec e_a.\vec\phi. \ee
The dominant walls are \beq && w_{-1} = \beta^3 - \beta^2\\ && w_0
= \beta^2 -\beta^1\\ && w_1 = \beta^1 -
\frac{\sqrt{2}}{2}\phi^{n-1}\\ && w_2 =
\frac{\sqrt{2}}{2}\,(\phi^{n-1} -\phi^{n-2}) \\ && \cdots \\ &&
w_{n-2} = \frac{\sqrt{2}}{2}\,(\phi^3 -\phi^2) \\ && w_{n-1}  =
\frac{\sqrt{2}}{2}\,(\phi^2 -\phi^1) \\ && w_{n} =
\sqrt{2}\,\phi^1 \end{eqnarray} The Cartan matrix is given by
$$  A = \left( \begin{array}{ccccccccc}
2 & -1 & 0 & 0 & \cdots & 0 & 0 & 0 &  0 \\
-1 & 2 & -1 & 0 & \cdots & 0 & 0 & 0 & 0\\
0 & -2 & 2 & -1 & \cdots & 0 & 0 & 0 & 0 \\
0 & 0 & -1 & 2 & \cdots & 0 & 0 & 0 & 0 \\
\vdots & \vdots & \vdots & \vdots & \ddots & \vdots & \vdots & \vdots
& \vdots \\
0 & 0 & 0 & 0 & \cdots & 2 & -1 & 0 & 0 \\
0 & 0 & 0 & 0 & \cdots & -1 & 2 & -1 & 0 \\
0 & 0 & 0 & 0 & \cdots & 0 & -1 & 2 & -2 \\
0 & 0 & 0 & 0 & \cdots & 0 & 0 & -1 & 2 \\
\end{array}
\right)
$$
The Dynkin diagram is given in figure 1. Again we recognize the Dynkin
diagram of the overextension  $C_n^{\wedge \wedge}$. It is hyperbolic
for $n\leq 4$.

\subsection{The $D_n$ sequence}
The oxidation end point Lagrangian in $D=n+2$ dimensions is the
following \be {\cal L}_D = R\star \unity - \star d\phi\wedge d\phi -
\frac{1}{2} \,e^{a\sqrt{2}\phi}\star dB\wedge dB\ee where $B$ is a
$2$-form and $a^2 = 8/n$.

Besides the curvature and the symmetry walls \beq && w_{-1} =
\beta^{n+1}-\beta^n,\;\;w_0 =
\beta^n -
\beta^{n-1}\\ && w_2= \beta^{n-1}-\beta^{n-2},..., w_{n-1} = \beta^2 -
\beta^1,\end{eqnarray} we get here the
$2$-form electric and magnetic walls \beq && w^E_{ij} = \beta^i +
\beta^j + a\frac{\sqrt{2}}{2}\phi\\ && w^M_{i_1...i_{n-2}} =
\beta^{i_1}+...+\beta^{i_{n-2}} -
a\frac{\sqrt{2}}{2}\phi.\end{eqnarray}
The dominant ones are
the $w_{-1}, w_0, w_2,...,w_{n-1}$ defined above and \beq && w_n =
\beta^1 +
\beta^2 + a\frac{\sqrt{2}}{2}\phi \\ && w_1 =
\beta^1+\beta^2+...+\beta^{n-2} -
a\frac{\sqrt{2}}{2}\phi.\end{eqnarray}

The Cartan matrix is given by
$$  A = \left( \begin{array}{ccccccccc}
2 & -1 & 0 & 0 & \cdots & 0 & 0 & 0 &  0 \\
-1 & 2 & 0 & -1 & \cdots & 0 & 0 & 0 & 0\\
0 & 0 & 2 & -1 & \cdots & 0 & 0 & 0 & 0 \\
0 & 0 & -1 & 2 & \cdots & 0 & 0 & 0 & 0 \\
\vdots & \vdots & \vdots & \vdots & \ddots & \vdots & \vdots & \vdots
& \vdots \\
0 & 0 & 0 & 0 & \cdots & 2 & -1 & 0 & 0 \\
0 & 0 & 0 & 0 & \cdots & -1 & 2 & -1 & -1 \\
0 & 0 & 0 & 0 & \cdots & 0 & -1 & 2 & 0 \\
0 & 0 & 0 & 0 & \cdots & 0 & -1 & 0 & 2 \\
\end{array}
\right)
$$

It is the generalized Cartan matrix of $D_n^{\wedge\wedge}=
DE_{n+2}$ which is known to be hyperbolic for
$n
\leq 8$. The corresponding Dynkin
diagram is given in figure 1.

For $n=8$, one gets the last hyperbolic algebra in this family,
namely $DE_{10} \equiv D_8^{\wedge\wedge}$, \cite{DH3}. For
$n=24$, which is the case relevant to the bosonic string, one gets
$D_{24}^{\wedge\wedge}$.

\begin{figure}[ht]
\centerline{\includegraphics[scale=0.8]{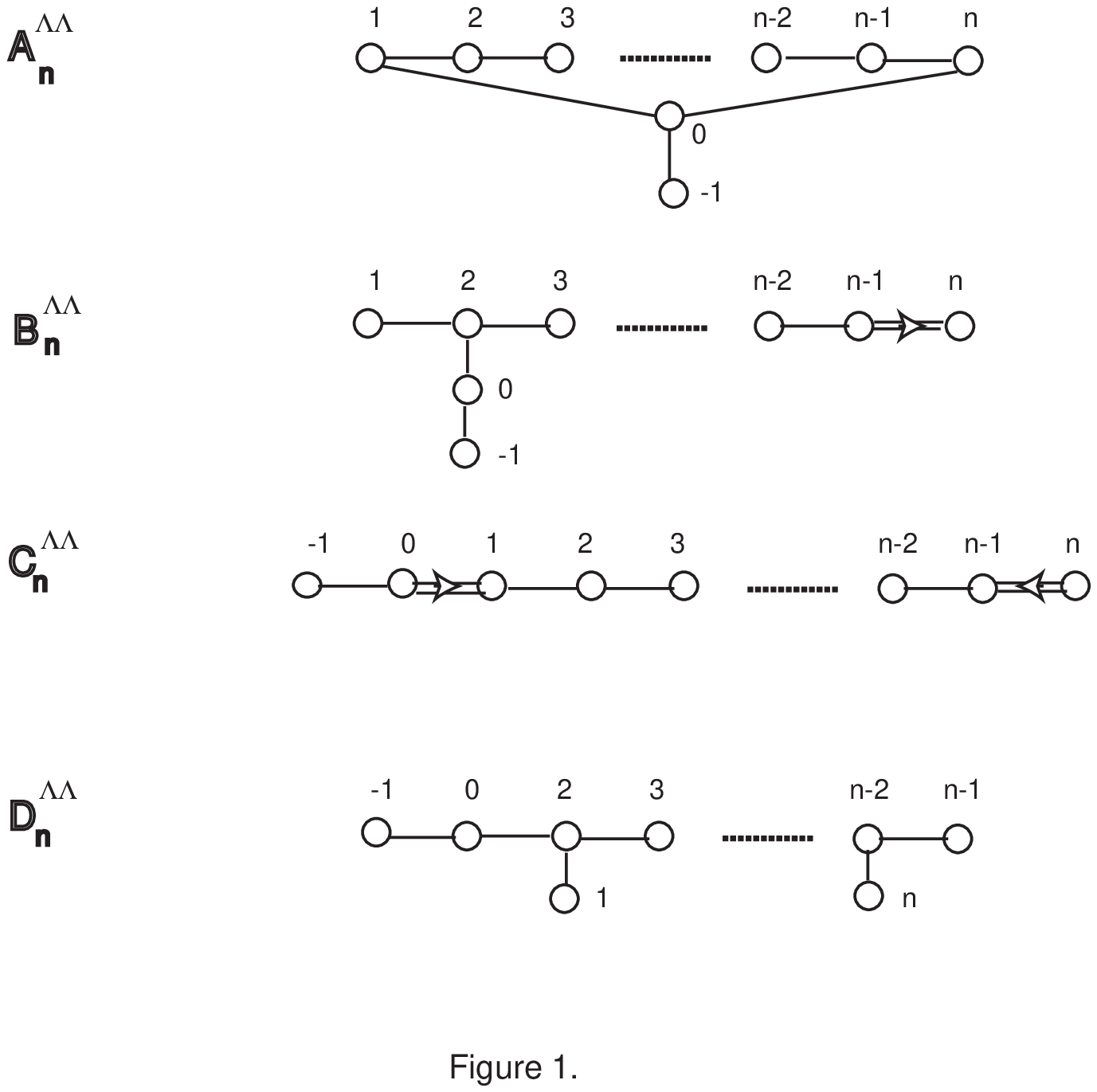}}
\end{figure}

\section{Models associated with exceptional groups}
\label{exceptional}
\setcounter{equation}{0}
\setcounter{theorem}{0} \setcounter{lemma}{0}

\subsection{The $G_2$-sequence}

The oxidation end point is the Einstein-Maxwell system in $D=5$
with an extra $FFA$ term \cite{CJLP}: \be {\cal L}_5 = R\star\unity -
\frac{1}{2}\star F\wedge F + \frac{1}{3\sqrt{3}} F\wedge F\wedge
A,\quad\quad F=dA.\ee

Besides the curvature and the symmetry walls, there are one-form
electric walls $w^E_i =\beta^i$ and magnetic walls $w^M_{ijk}=
\beta^i + \beta^j + \beta^k,\quad i<j<k $.  The dominant walls are \beq
&& w_{-1} =
\beta^4 -\beta^3 \\ && w_0 = \beta^3 - \beta^2 \\ && w_1 = \beta^2 -
\beta^1\\ && w_2 = \beta^1.\end{eqnarray} The Cartan
matrix reads

$$  A = \left( \begin{array}{cccc}
2 & -1 & 0 & 0\\
-1 & 2 & -1 & 0 \\
0  & -1 &  2 & -1\\
0 & 0  & -3  & 2 \\
\end{array} \right)
$$
which is the generalized Cartan matrix of $G_2^{\wedge\wedge}$. Its
Dynkin diagram is given in figure 2; the algebra is hyperbolic
\cite{DHJN}.  See also \cite{Mizo} for the relevance of
$G_2^{\wedge\wedge}$ to this system.

\subsection{The $F_4$-sequence}
The oxidation end point of the $F_4$ sequence is a $D=6$
dimensional theory containing the metric, a dilaton ($\phi)$, an
axion ($\chi)$, two one-forms $(A^{\pm})$, a two-form $(B)$ and a
self-dual $3$-form field strength $(G)$ \cite{CJLP}. The
Lagrangian is given by \beq {\cal L}_6 = && R\star \unity - \star
d\phi\wedge d\phi -\frac{1}{2} e^{2\phi}\star d\chi\wedge d\chi -
\frac{1}{2} e^{-2\phi} \star H\wedge H
\\ && - \frac{1}{2} \star G \wedge G -\frac{1}{2}   e^{\phi}
\star F^{+}\wedge F^{+} -\frac{1}{2}    e^{-\phi} \star
F^{-}\wedge F^{-} \\ && -\frac{1}{\sqrt{2}}\chi \,H\wedge G
-\frac{1}{2} A^+\wedge F^+\wedge H -\frac{1}{2} A^+\wedge
F^-\wedge G.\end{eqnarray} The field strengths are given in terms
of potentials as follows: \beq && F^+ = d A^+ + \frac{1}{\sqrt{2}}
\chi\,dA^-\\ && F^- = dA^-\\ && H = dB + \frac{1}{2} A^-\wedge
dA^- \\ && G = d C - \frac{1}{\sqrt{2}}\,\chi\,H - \frac{1}{2}
A^+\wedge d A^-.\end{eqnarray} Besides the curvature and the
symmetry walls we get here
\begin{enumerate}
\item {Electric walls} \beq && w^{E,\chi} = \phi\\ && w^{E,A^+}_i
= \beta^i + \frac{\phi}{2} \\ && w^{E,A^-}_i = \beta^i -
\frac{\phi}{2} \\ && w^{E,B}_{ij} = \beta^i + \beta^j - \phi,\quad
i<j \\ && w^{E,C}_{ij} = \beta^i + \beta^j,\quad i<j\end{eqnarray}
\item {Magnetic walls} \beq && w^{M,\chi}_{ijk\ell} = \beta^i +
\beta^j + \beta^k + \beta^\ell -\phi,\quad i<j<k<\ell\\ &&
w^{M,A^+}_{ijk} = \beta^i + \beta^j + \beta^k
-\frac{\phi}{2},\quad i<j<k \\ && w^{M,A^-}_{ijk} = \beta^i
+\beta^j + \beta^k + \frac{\phi}{2},\quad i<j<k \\ && w^{M,B}_{i}
= \beta^i + \phi\\ && w^{M,C}_{i} = \beta^i
\end{eqnarray}
\end{enumerate} The dominant walls are \beq && w_{-1} = \beta^5 -
\beta^4\\ && w_0 =
\beta^4 -
\beta^3
\\ && w_1 = \beta^3 -\beta^2 \\ && w_2 = \beta^2 - \beta^1\\ && w_3 =
\beta^1 - \frac{\phi}{2} \\ && w_4 = \phi.\end{eqnarray}

The Cartan matrix is given by $$  A = \left( \begin{array}{cccccc}
2 & -1 & 0 & 0 & 0 & 0 \\
-1 & 2 & -1 & 0 & 0 & 0 \\
0  & -1 &  2 & -1 & 0 & 0 \\
0 & 0  & -1  & 2 & -1 & 0 \\
0 & 0 & 0  &  -2 & 2 & -1 \\
0 & 0 & 0  & 0  & -1 & 2 \\
  \end{array}
\right)
$$ This is the generalized Cartan matrix of $F_4^{\wedge\wedge}$; its
Dynkin diagram is given in figure 2.  The overextended algebra is
hyperbolic.

\subsection{The $E_6$-sequence}
The oxidation end point of the $E_6$ sequence is $D=8$ and the
associated theory is the smallest obtainable as a truncation of
maximal supergravity in which the $3$-form potential is retained.
It comprises the metric, a dilaton and an axion,$\chi$, together
with the $3$-form, $C$ \cite{CJLP}. The $8$-dimensional Lagrangian
is given by \be {\cal L}_8 = R\star \unity - \star d\phi\wedge d\phi
-\frac{1}{2} e^{2\sqrt{2}\phi}\star d\chi\wedge d\chi -
\frac{1}{2}e^{-\sqrt{2}\phi}\star G\wedge G + \chi\,G\wedge G,\ee
where $G=dC$. In addition to the gravitational and symmetry walls,
there are the electric and magnetic walls of the scalar \be
w^{E,\chi} = \sqrt{2}\phi\quad\hbox{and}\quad
w^{M,\chi}_{i_1...i_6} = \beta^{i_1}+...+\beta^{i_6}+
\sqrt{2}\phi\ee and those coming from the $3$-form \beq &&
w^{E,C}_{ijk} = \beta^i + \beta^j + \beta^k -
\frac{\sqrt{2}}{2}\phi \\ && w^{M,C}_{ijk} = \beta^i + \beta^j +
 \beta^k +
\frac{\sqrt{2}}{2}\phi.\end{eqnarray}
The dominant walls are the
symmetry walls \beq && w_{-1} = \beta^7 -\beta^6,\;\;w_0= \beta^6
-\beta^5 \\ && w_1 = \beta^2-\beta^1,\;\; w_2 =
\beta^3-\beta^2,\;\;w_3 = \beta^4 -\beta^3,\;\;w_6 = \beta^5
-\beta^4 \end{eqnarray} and
\beq && w_4 =
\beta^1+\beta^2+\beta^3 - \frac{\sqrt{2}}{2}\phi \\ && w_{5} =
\sqrt{2}\phi.\end{eqnarray}

The Cartan matrix is given by
 $$  A = \left( \begin{array}{cccccccc}
2 & -1 & 0 & 0 & 0 & 0 & 0 & 0\\
-1 & 2 & 0 & 0 & 0 & 0 & 0 & -1\\
0  & 0 & 2 &  -1 & 0 & 0 & 0 & 0 \\
0 & 0  & -1  & 2 & -1 & 0 & 0 & 0 \\
0 & 0 & 0  &  -1 & 2 & -1 & 0 & -1\\
0 & 0 & 0  & 0  & -1 & 2 & -1 & 0\\
0 & 0 & 0 & 0 & 0 & -1 & 2 & 0 \\
0 & -1 & 0 & 0 & -1 & 0 & 0 & 2\\  \end{array}
\right) $$

It is the generalized Cartan matrix of the
hyperbolic algebra $E_6^{\wedge\wedge}$.

The Dynkin diagram is given in figure 2.

\subsection{The $E_7$-sequence}
This sequence is obtained as a consistent (albeit non
supersymmetric) truncation of $D=9$ maximal supergravity to the
theory whose bosonic sector comprises the metric, a dilaton, a
$1$-form, $A$, and a $3$-form potential $C$ \cite{CJLP}. The
Lagrangian reads as \beq {\cal L}_9 = && R\star \unity - \star
d\phi\wedge d\phi -\frac{1}{2}
e^{\frac{2\sqrt{2}}{\sqrt{7}}\phi}\star dC\wedge dC
\\ && -\frac{1}{2} e^{-\frac{4\sqrt{2}}{\sqrt{7}}\phi}\star
dA\wedge dA -\frac{1}{2} dC\wedge dC\wedge
A.\nonumber\end{eqnarray}

Besides the curvature and the symmetry walls, we have here the
electric and magnetic malls of the $1$-form \beq && w^{E,A}_i = \beta^i
-\frac{2\sqrt{2}}{\sqrt{7}}\phi\\ && w^{M,A}_{i_1...i_6}=
\beta^{i_1}+...\beta^{i_6} +
\frac{2\sqrt{2}}{\sqrt{7}}\phi\end{eqnarray} and the corresponding
walls of the $3$-form \beq && w^{E,C}_{ijk} = \beta^i + \beta^j+\beta^k
+\frac{\sqrt{2}}{\sqrt{7}}\phi\\ && w^{M,C}_{ijk\ell}=
\beta^{i}+\beta^j+\beta^k+\beta^{\ell} -
\frac{\sqrt{2}}{\sqrt{7}}\phi\end{eqnarray}
The dominant walls are
the symmetry walls \be w_{-1} = \beta^8-\beta^7, w_0=
\beta^7-\beta^6,  w_1= \beta^6-\beta^5,...,\, w_5 = \beta^2 -
\beta^1\ee and
\beq  && w_6 = \beta^1
-\frac{2\sqrt{2}}{\sqrt{7}}\phi \\ && w_7 =
\beta^1+\beta^2+\beta^3 +
\frac{\sqrt{2}}{\sqrt{7}}\phi.\end{eqnarray}
The Cartan matrix reads
$$  A = \left( \begin{array}{ccccccccc}
2 & -1 & 0 & 0 & 0 & 0 & 0 & 0 &  0 \\
-1 & 2 & -1 & 0 & 0 & 0 & 0 & 0 & 0\\
0 & -1 & 2 & -1 & 0 & 0 & 0 & 0 & 0 \\
0 & 0 & -1 & 2 & -1 & 0 & 0 & 0 & 0 \\
0 & 0 & 0 & -1 & 2 & -1 & 0 & 0
& -1 \\
0 & 0 & 0 & 0 & -1 & 2 & -1 & 0 & 0 \\
0 & 0 & 0 & 0 & 0 & -1 & 2 & -1 & 0 \\
0 & 0 & 0 & 0 & 0 & 0 & -1 & 2 & 0 \\
0 & 0 & 0 & 0 & -1 & 0 & 0 & 0 & 2\\
\end{array}
\right)
$$

The Dynkin diagram is given in figure 2; again it is the overextension
$E_7^{\wedge\wedge}$, which is hyperbolic.

In fact, one can view the above Lagrangian as the reduction to $9$
spacetime dimensions of the type IIB supergravity theory in which
one keeps only the metric and the chiral $4$-form \cite{CJLP}.
{}From the point of view of getting the correct walls, the
self-duality condition on the field strength is actually not
necessary: without the chirality condition, each $4$-form walls
would simply appear twice (once as electric, once as magnetic wall).

\subsection{The $E_8$-sequence}
The oxidation end point is $D=11$-dimensional supergravity whose
bosonic sector is given by \be {\cal L}_{11} = R\star \unity
-\frac{1}{2}\star dC\wedge dC -\frac{1}{6} dC\wedge dC\wedge C\ee $C$
is a $3$-form.

Besides the curvature and the symmetry walls, we have here the
electric and magnetic walls of the $3$-form which read as \beq
&& w^{E}_{ijk} = \beta^i + \beta^j + \beta^k \\ && w^{M}_{i_1...i_6} =
\beta^{i_1}+...+\beta^{i_6}.\end{eqnarray}
The dominant walls are
the symmetry walls \be w_{-1} = \beta^{10}-\beta^9, w_0=
\beta^9-\beta^8, w_1= \beta^8-\beta^7,..., w_7 = \beta^2 - \beta^1\ee
and
\be w_{8} =
\beta^1+\beta^2+\beta^3.\ee
The Cartan matrix is
$$  A = \left( \begin{array}{cccccccccc}
2 & -1 & 0 & 0 & 0 & 0 & 0 & 0 &  0 & 0\\
-1 & 2 & -1 & 0 & 0 & 0 & 0 & 0 & 0 & 0\\
0 & -1 & 2 & -1 & 0 & 0 & 0 & 0 & 0 & 0\\
0 & 0 & -1 & 2 & -1 & 0 & 0 & 0 & 0 & 0\\
0 & 0 & 0 & -1 & 2 & -1 & 0 & 0
& -1 & 0\\
0 & 0 & 0 & 0 & -1 & 2 & -1 & 0 & 0 & 0\\
0 & 0 & 0 & 0 & 0 & -1 & 2 & 0 & 0 & -1\\
0 & 0 & 0 & 0 & 0 & 0 & -1 & 2 & -1 & 0\\
0 & 0 & 0 & 0 & 0 & 0 & 0 & -1 & 2 & 0\\
0 & 0 & 0 & 0 & 0 & 0 & -1 & 0 & 0 & 2\\
\end{array}
\right)
$$

As pointed out in \cite{DH3} this is the Cartan matrix of the
overextension $E_8^{\wedge\wedge}$, better known as $E_{10}$.
As shown in that paper, it is also the Cartan matrix relevant
to type IIA supergravity in ten
dimensions (dimensional reduction) as well as type IIB.  In fact,
massive type IIA supergravity in ten dimensions \cite{Romans} is
also described by the same billiard, as one can easily verify by
using the formulation of \cite{BdRGPT}: the new wall associated
with the mass term can be expressed as a linear combination with
positive (integer) coefficients of the dominant walls and is thus
subdominant (note that it has squared norm equal to $2$).

The Dynkin diagram is given in figure 2; it is hyperbolic.
The relevance of $E_{10}$ in the supergravity context was
first conjectured in \cite{julia80}.

\begin{figure}[ht]
\centerline{\includegraphics[scale=0.8]{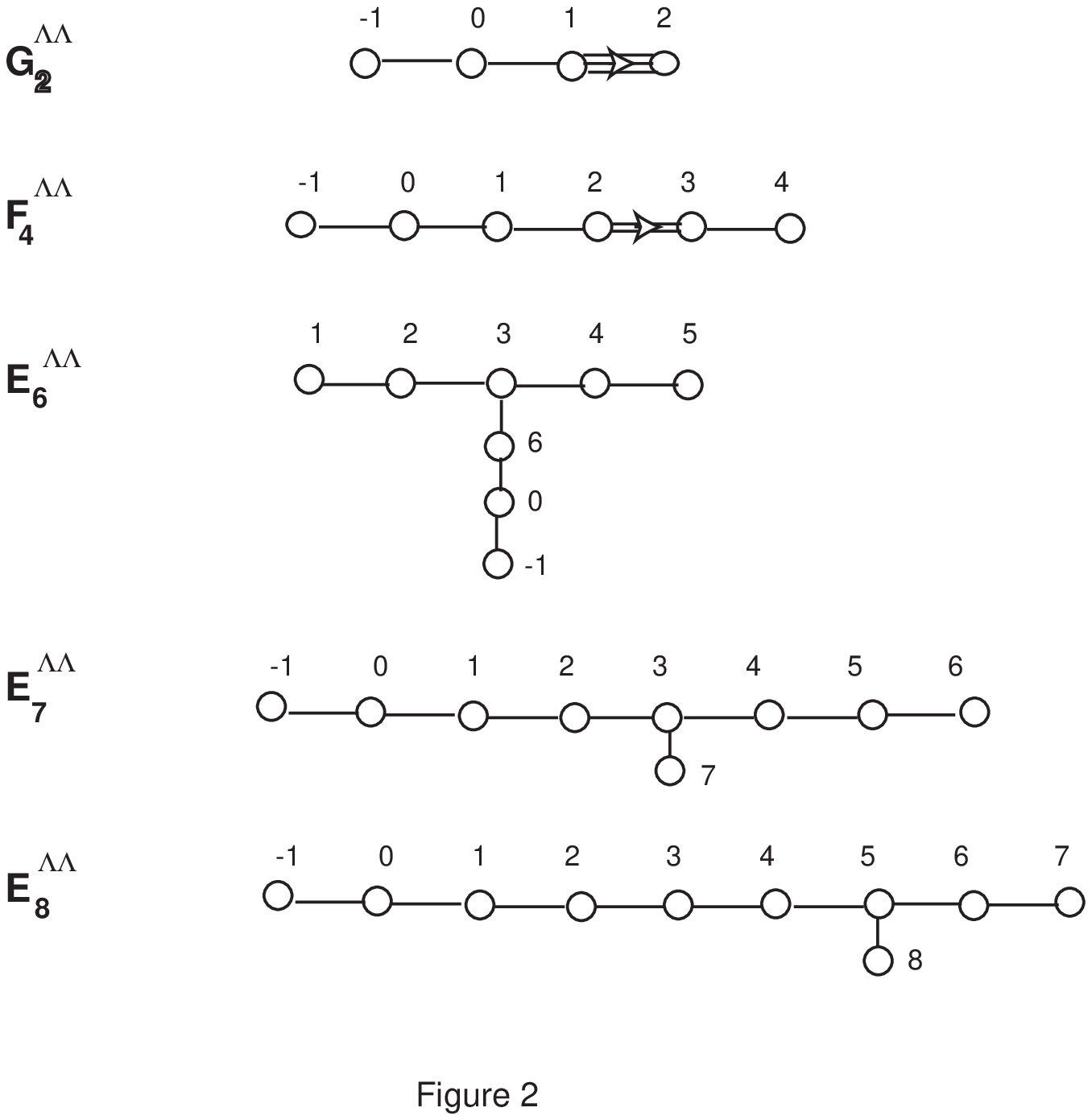}}
\end{figure}

\break

\section{Conclusions}
\setcounter{equation}{0} \setcounter{theorem}{0}
\setcounter{lemma}{0}

In this paper, we have provided models that realize the canonical
Lorentzian extensions, or overextensions, of all the
finite-dimensional Lie algebras, in the sense that their
asymptotic dynamics in the vicinity of a spacelike singularity is
a billiard motion in the fundamental Weyl chamber of the
corresponding Kac-Moody algebras.  We have also shown that the
computation can be done at any point along the oxidation sequence.

The explicit proof given above of the appearance of the
overextensions was achieved by a case by case analysis, and was
based on the calculation of the Cartan matrix of the billiard as
it appears in the highest possible dimension (i.e. at the upper
edge of the oxidation sequence). As we said, another natural
dimension in which to compute the Cartan matrix of the billiard is
the lower edge of the oxidation sequence, namely in $D=3$
spacetime dimensions, where the scalar Lagrangian is the one of a
$G/H$ coset model. In this concluding section, we complete our
work by indicating how the calculation goes in $D=3$. We are going
to see that the calculation in this dimension has the advantage of
streamlining the essential structural elements which are at the
origin of the appearance of overextensions. This allows one to
better understand why the overextension of ${\cal G}$
systematically appears.

In $D=3$ dimensions, each Lagrangian model of \cite{BGM,CJLP}
exhibits  the following general structure \be {\cal L}_3 = R\star
\unity - \sum_{\alpha =1}^r \star d\phi^\alpha\wedge d\phi^\alpha
-\frac{1}{2}\sum_{A} e^{2\,\vec \sigma_A. \vec \phi} \star (d
\chi^A + \cdots)\wedge (d\chi^A + \cdots) \label{structure}\ee
where the dilatons $\phi^\a\; (\a=1,...,r) $ are associated with a
Cartan subalgebra of the simple Lie algebra ${\cal G}$ of rank $r$
and the scalar axions $\chi^A$ are associated with all the
positive roots $\vec \sigma_A$ of ${\cal G}$. In this Lagrangian,
the ellipsis $\cdots$ result from modifications of the
``curvatures". Such modifications are essential for ensuring the
presence of an exact $G$-symmetry. However, they are not important
for computing the Cartan matrix of the corresponding cosmological
billiard. The notation $\vec \sigma_A. \vec \phi$ stands for
$$ \vec \sigma_A. \vec \phi \equiv \sum_{\a = 1}^{r}
\sigma_A^\a \phi^\a.$$ It could also be denoted $\sigma_A( \phi)$
to emphasize that $\sigma_A$ is a linear form of the $\phi$'s. A
key feature for deriving our result is the fact that the norms
of the roots in the Lagrangians of \cite{CJLP} are such that the
long roots (and in particular the highest root $\theta$) have
squared length equal to $2$.  Indeed, if we were to focus only on
the $G$-symmetry in $D=3$ dimensions, one could multiply the
scalar part of the Lagrangian (\ref{structure}) by an arbitrary
number $k \not=0$ without changing its $G$-invariance.  This would
multiply the norms of all the non-gravitational 
roots by $1/k$.  However, such
models in which one changes the relative normalization of the
Einstein-Hilbert action with respect to the scalar coset action
cannot be oxidized to dimensions $D>3$ \cite{CJLP}.  Our
computation below shows that the correct relative normalization for
"oxidability" to higher dimensions is also crucial for getting the
overextension.

Note that, in $D=3$ all the non-gravitational walls are due to the
energy associated to scalar fields: the axions $\chi^A$. (We
recall that, in $D=3$, one can use dualisations to replace the
other fields, i.e. the one-forms if any, by scalars). Still, we
must take into account both the electric walls of the axions
(linked to the time derivatives of $\chi^A$) and their magnetic
walls (linked to their space derivatives). The axion electric
walls depend only on the dilatons and read \be w^{E,\chi^A} = \vec
\sigma_A.\vec \phi.\ee The axion magnetic walls are \be
w^{M,\chi^A}_i= \beta^i - \vec \sigma_A.\vec \phi. \label{mag3}\ee
Finally, since $d=2$, there is only one symmetry wall \be w^S =
\beta^2 - \beta^1.\ee Among these, the dominant walls are
\begin{enumerate}
\item{The
 electric walls} built from a set of
simple roots $\vec\sigma_\a$ ($\a = 1, \cdots, r$) and written as
\be w_\a = \vec \sigma_\a.\vec\phi\ee Indeed, all the other
electric walls will be, by definition, combinations of such simple
walls with positive coefficients. \item{The magnetic wall}
associated (in view of (\ref{mag3}))
 to the lowest $\b^i$ (i.e.
$\b^1$, because $\b^2 = \b^1 + w^S)$, and to the highest root of ${\cal G}$.
 This yields as dominant
magnetic wall
\be w_0 =
 \beta^1 - \vec\theta.\vec\phi \label{domin}\ee where
$\vec \theta = \sum_\alpha n_\alpha \vec \s_\alpha $ denotes the
highest root of ${\cal G}$.
\item{The symmetry wall}: \be w_{-1} = w^S = \beta^2 - \beta^1.\ee
\end{enumerate}

Let us now notice that the two linear forms $-\beta^1$ and
$\beta^2$ have zero norm and scalar product equal to $+1$.
Moreover, they are, by definition orthogonal (in the sense of the
co-metric (\ref{Gmn})) to all the $\phi$-dependent forms.
Therefore they play exactly the r\^ole of the additional null
roots $u_1$ and $u_2$ introduced in section \ref{overex}, and can
be identified with these. Explicitly: $-\beta^1 \to u_2$ and
$\beta^2 \to u_1$. One then sees that the dominant wall forms
above can be identified with the simple roots of the overextension
${\cal G}^{\wedge\wedge}$ of ${\cal G}$. More precisely, the
relevant electric wall-forms are the simple roots of ${\cal G}$,
the magnetic wall-form  is the root which does the affine
extension ${\cal G}^\wedge$ and the symmetry wall-form  is the one
responsible for the overextension. This completes our second
(general) proof of the appearance of the overextension   ${\cal
G}^{\wedge\wedge}$ in the cosmological billiards.

The magnetic wall (\ref{domin}) is in fact also a symmetry wall in
the last dimension of the oxidation sequence; thus, as all
gravitational walls, it has squared norm equal to two. Since
$-\beta^1$ has zero norm, this explains the origin of the
normalization of the highest root $\vec \theta$ in our context.

{}Finally, we note that the invariance of the billiard under
dimensional reduction may help in conjecturing which symmetry
group arises in the toroidal compactification down to three
dimensions of a theory given in $D>3$ dimensions, without having
to actually carry out this reduction.

\section*{Acknowledgements}
M.H. is grateful to Bernard Julia for informative discussions.The
work of S.d.B., M.H. and C.S. is supported in part by the
``Actions de Recherche Concert{\'e}es" of the ``Direction de la
Recherche Scientifique - Communaut{\'e} Fran{\c c}aise de
Belgique", by a ``P\^ole d'Attraction Interuniversitaire"
(Belgium), by IISN-Belgium (convention 4.4505.86), by Proyectos
FONDECYT 1970151 and 7960001 (Chile) and by the European
Commission RTN programme HPRN-CT-00131, in which they are
associated to K. U. Leuven.

\end{document}